\documentclass[review]{elsarticle}
\usepackage{amsmath,graphicx,dsfont,ctable,amssymb,amsfonts,mathtools} 
\usepackage{algorithm}
\usepackage{algorithmic}
\usepackage{dblfloatfix}
\usepackage{amsfonts}
\usepackage{caption}
\usepackage{hyperref}
\usepackage{makecell}
\usepackage{perpage} 
\usepackage{hyperref}

\journal{Ultrasonics}
\begin{document}

\begin{frontmatter}

\title{Plane-Wave Ultrasound Beamforming: A Deep Learning Approach}

\author{Sobhan Goudarzi\corref{mycorrespondingauthor}}
\cortext[mycorrespondingauthor]{Corresponding author}
\ead{sobhan.goudarzi@concordia.ca}
\author{Hassan Rivaz}
\address{Department of Electrical and Computer Engineering, Concordia University, Montreal, QC, Canada.}

\begin{abstract}
Medical ultrasound provides images which are the spatial map of the tissue echogenicity. Unfortunately, an ultrasound image is a low-quality version of the expected Tissue Reflectivity Function (TRF) mainly due to the non-ideal Point Spread Function (PSF) of the imaging system. This paper presents a novel beamforming approach based on deep learning to get closer to the ideal PSF in Plane-Wave Imaging (PWI). The proposed approach is designed to reconstruct the desired TRF from echo traces acquired by transducer elements using only a single plane-wave transmission. In this approach, first, an ideal model for the TRF is introduced by setting the imaging PSF as a sharp Gaussian function. Then, a mapping function between the pre-beamformed Radio-Frequency (RF) channel data and the proposed TRF is constructed using deep learning. Network architecture contains multi-resolution decomposition and reconstruction using wavelet transform for effective recovery of high-frequency content of the desired TRF. Inspired by curriculum learning, we exploit step by step training from coarse (mean square error) to fine ($\ell_{0.2}$) loss functions. The proposed method is trained on a large number of simulation ultrasound data with the ground-truth echogenicity map extracted from real photographic images. The performance of the trained network is evaluated on the publicly available simulation and \textit{in vivo} test data without any further fine-tuning. Simulation test results confirm that the proposed method reconstructs images with a high quality in terms of resolution and contrast, which are also visually similar to the proposed ground-truth image. Furthermore, \textit{in vivo} results show that the trained mapping function preserves its performance in the new domain. Therefore, the proposed approach maintains high resolution, contrast, and framerate simultaneously.
\end{abstract}


\end{frontmatter}

\section{Introduction}
\label{sec:sec1}
Plane-Wave Imaging (PWI) is of the highest framerate among different ultrasound imaging techniques. In this method, the whole medium is insonified in a single shot using a plane-wave produced by firing all transducer's elements simultaneously. Then, all piezoelectric elements record the backscattered signals from the medium. Therefore, the framerate can reach several thousand frames per second since it is only limited by the depth of imaging and the speed of sound. Each element's output gives a low-quality spatial map of the target echogenicity, and the final image is the result of combining all elements' outputs.\par
\begin{figure}[t!]
	\centering
	\centerline{\includegraphics[width=8.5cm]{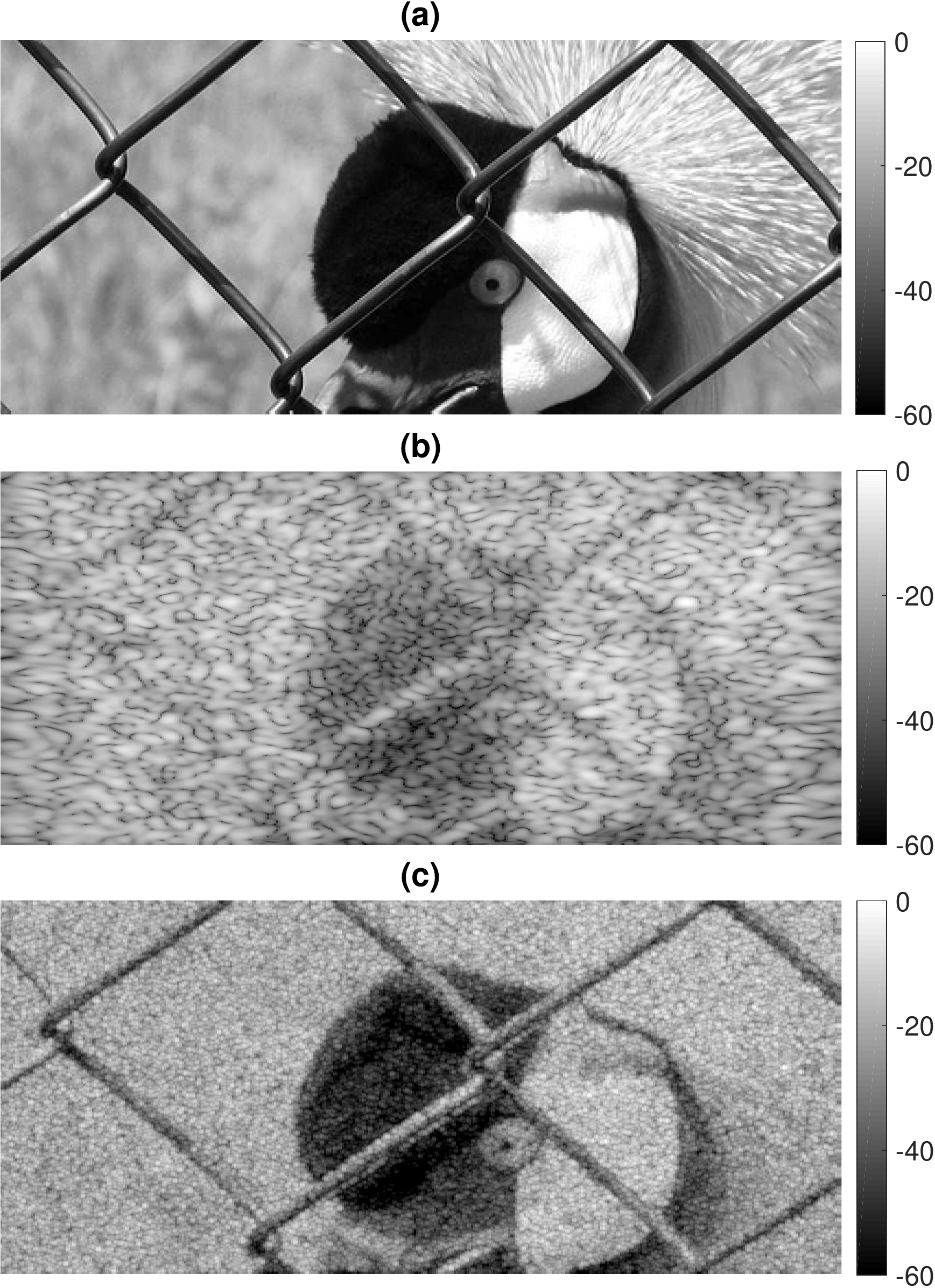}}
	\caption{Demonstration of the PSF effect on the quality of ultrasound imaging. (a) Ground-truth echogenicity map extracted from a real photographic image from an ImageNet~\cite{ILSVRC15} validation set. (b) Simulated ultrasound image from a single $0^o$ plane-wave transmission. (c) Desired ultrasound image reconstructed using the proposed ideal PSF.}
	\label{fig:fig5}
\end{figure}

The resulting image, however, is of low quality mainly due to the non-ideal Point Spread Function (PSF) of the imaging system. Fig.~\ref{fig:fig5} compares images created with PWI to an image created with a sharp Gaussian PSF as an ideal ultrasound image. The common approach in PWI provides a low quality version of the expected Tissue Reflectivity Function (TRF) shown in Fig.~\ref{fig:fig5}(a). There are several reasons that render the PWI image of low quality such as unfocused transmission, limited frequency response of the piezoelectric elements, and limitation of transducer design. Some of these reasons are intrinsic physical limitations which exist among all ultrasound imaging techniques. Given the very high framerate of PWI, improving its image quality is an active field of research.\par
Coherent Plane-Wave Compounding (CPWC) is a skillful solution to the problem of reduction in image quality due to the unfocused transmission~\cite{4816058}. This solution, however, comes at the expense of reduction in framerate, leading to a trade-off between image quality and framerate. In addition, this solution leads to motion blurs in applications such as shear-wave imaging and echocardiography. Therefore, a growing body of research has focused on transcending this limitation~\cite{7728908,9251434,9194075}.\par
Deep learning has shown an immense potential for different applications in medical imaging such as segmentation, reconstruction and classification~\cite{zhou2020review,8808885,9264709,9173689}. It is an efficient tool for solving ill-posed problems with non-linear mapping functions between high dimensional input-output pairs. Deep learning, however, suffers from three main limitations in medical imaging applications including the scarcity of training data, lack of ground-truth labelled data, and domain shift between training and test data~\cite{10.5555/1462129}.\par   
Considering the aforementioned limitations, deep learning is exploited herein for PWI beamforming. Our goal is to extract the best possible image quality from the Radio-Frequency (RF) channel data. To do so, an ideal PSF (sharp Gaussian function) is proposed for ultrasound PWI, and the result of convolution between the TRF and the proposed PSF is considered as ground-truth (Fig.~\ref{fig:fig5}(c)). Then, deep learning is used to construct a mapping function between the RF channel data and the proposed ideal image. Herein, the necessity of having a large number of training data is resolved by simulating pre-beamformed ultrasound RF data from real photographic images. Finally, the generalization performance of the trained beamforming network is evaluated on publicly available simulation as well as \textit{in vivo} test datasets. The main contributions of the paper can be summarized as follows:
\begin{enumerate}
	\item To obtain the optimal image quality in PWI, a sharp Gaussian mask is proposed as the desired PSF, and used for the simulation of the training data.
	\item The proposed beamforming network architecture contains multi-resolution decomposition and reconstruction using wavelet transform, with the best match mother wavelet to the RF data, for effective recovery of high-frequency content of the desired TRF. 
	\item Given the high complexity of the deep representation, we exploit curriculum learning by successively employing more challenging loss functions.  
	\item The proposed approach preserves its performance on the unseen test data without any further fine-tuning.  
	\item In addition to reducing side- and grating-lobes, the proposed method improves the axial resolution. Traditional beamforming methods do not improve the axial resolution, which can only be improved with increasing the center frequency and subsequently reducing the penetration depth.
\end{enumerate}

\subsection{Related work}
\label{sec:sec11}
The main focus of the current paper is the application of deep learning in ultrasound imaging and specifically in image reconstruction and beamforming. Therefore, only related important researches, among a large number of recent literatures, are reviewed. Luchies~\textit{et al.} proposed an approach for ultrasound beamforming based on Deep Neural Network (DNN)~\cite{8302520}. More specifically, a distinct DNN is trained on In-phase/Quadrature (IQ) data of each frequency sub-band to suppress off-axis scattering. They used simulated point target responses for training. Subsequently, the authors extended this approach by using simulated and phantom anechoic cysts for training~\cite{9116815}. The reconstruction of B-mode ultrasound images from sub-sampled RF data using deep learning was investigated in~\cite{8432500}. Convolutional Neural Networks (CNNs) have also been used to reconstruct the image quality of CPWC from a single transmission~\cite{8436013}. Hyun~\textit{et al.} made use of Fully CNN (FCNN) for reconstructing despeckled ultrasound images~\cite{8663450}. An autoencoder network structure, with a single encoder and two decoders, was used to simultaneously obtain the segmentation map and B-mode ultrasound image from a single plane-wave transmission~\cite{9090910}. Deep learning was employed as an adaptive ultrasound beamforming technique in~\cite{9138451,9025198}. Goudarzi~\textit{et al.} proposed an approach based on Generative Adversarial Neural Network (GAN) to achieve the quality of multi-focus ultrasound imaging using only a single focused transmission~\cite{9178454}. High quality image reconstruction of diverging-wave ultrasound imaging from a small number of transmissions based on CNNs is proposed in~\cite{9057682}.\par
Recently, the Challenge on Ultrasound Beamforming with Deep Learning (CUBDL) was held in conjunction with the 2020 IEEE International Ultrasonics Symposium (IUS)~\cite{9251434}. We have successfully participated in this challenge and proposed a general approach for ultrasound beamforming using deep learning~\cite{9251565}. More specifically, MobileNetV2~\cite{Sandler_2018_CVPR} structure has been adapted to train a model that mimics Minimum Variance Beamforming (MVB). In terms of image quality, our method was ranked first. Overall, considering both image quality and network size, our method was jointly ranked first with another submission~\cite{9251322}. We compare the new proposed beamforming technique to the one we proposed in CUBDL. \par
The rest of the paper is organized as follows. First, the proposed method is introduced in Section~\ref{sec:sec2}. Details of datasets, network training, and evaluation metrics are explained in Section~\ref{sec:sec3}. Results are presented in Section~\ref{sec:sec4} and discussed in Section~\ref{sec:sec5}. Finally, the paper is concluded in Section~\ref{sec:sec6}. 
\section{Methods}
\label{sec:sec2}
\begin{figure}[b!]
	\centering
	\centerline{\includegraphics[width=7.5cm]{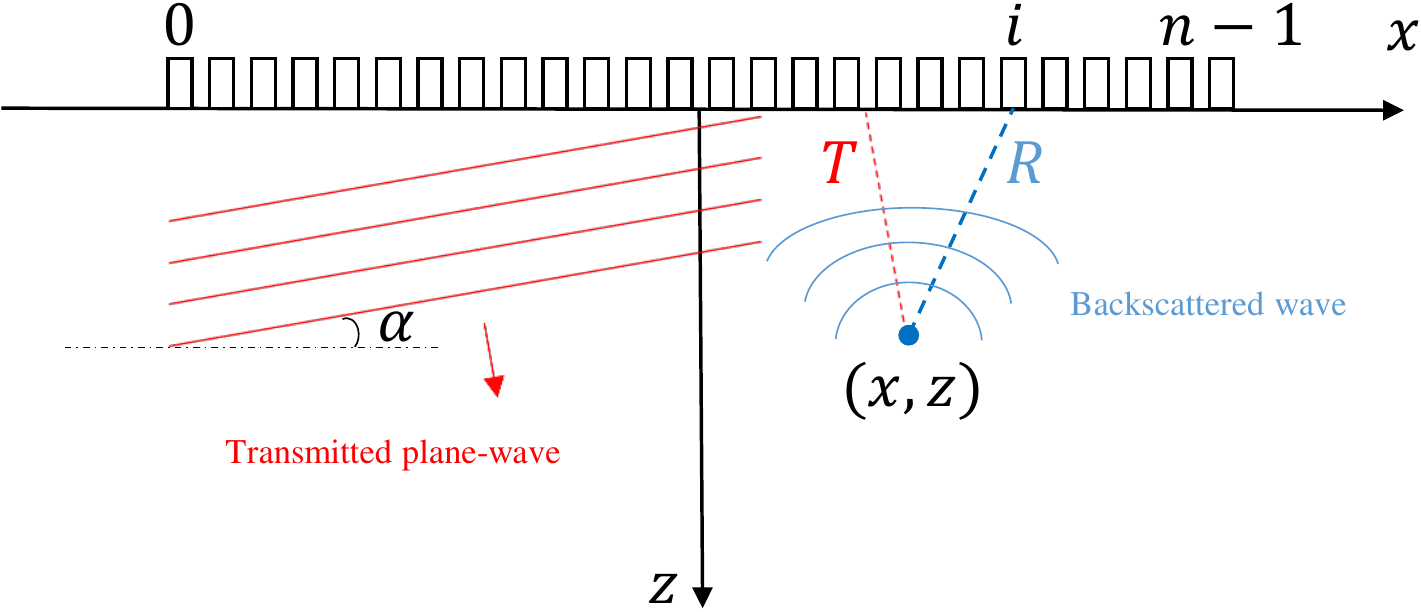}}
	\caption{Geometrical illustration of the PWI. The steering angle of the incidence wave is denoted by $\alpha$.}
	\label{fig:fig7}
\end{figure}
\subsection{Plane-wave beamforming}
\label{sec:sec21}
In this section, an overview of PWI using classical Delay-And-Sum (DAS) beamforming is provided. More details regarding the PWI as well as CPWC techniques can be found in~\cite{4816058}. 
Let us assume a linear probe wherein $n$ crystal elements are distributed on the $x$-axis and transmit ultrasound waves into the medium along the positive $z$-axis  (Fig.~\ref{fig:fig7}). Consider that the medium with a sound speed of $c$ is insonified with a plane-wave with angle $\alpha$. The backscattered waves recorded by crystal element $i$ located at $x_i$ is denoted by $\mathbf{h}_i(t)$. It has been shown ~\cite{4816058} that the transmission distance $d_t$ from the origin of the transmitted plane-wave to an arbitrary point $(x,z)$ in the Region-Of-Interest (ROI) equals $zcos(\alpha)+xsin(\alpha)$, and the receiving distance $d_r$ from $(x,z)$ to the location of crystal element $i$ equals $\sqrt{ (x-x_i)^{2}+z^2 }$.\par 
The spatial map of scatterers can be discretized to generate image pixels, wherein each pixel corresponds to an averaged TRF over the extent of the pixel. We need to extract the RF data corresponding to each pixel of the final image from the output
of each crystal element. Therefore, if we define $R_i$ as a matrix containing the RF data recorded by crystal element $i$ corresponding to each pixel $(x,z)$ in the ROI, its elements can be found by applying the associated propagation delay to $\mathbf{h}_i(t)$ as follows (hereafter, capital and bold font variables represent matrices and vectors, respectively):
\begin{equation} 
	\label{eq:4}
	\tau(x,z) =\frac{d_t+d_r}{c}  \Longrightarrow  R_i(x,z) = \mathbf{h}_i(\tau(x,z)) ,
\end{equation}
As seen in Eq.~(\ref{eq:4}), each probe element gives one RF matrix of the ROI. Finally, the beamformed RF matrix $S$ is the result of information fusion among different crystal elements, and each element of $S(x,z)$ can be obtained through the following weighted summation:
\begin{equation} 
	\label{eq:5}
	S(x,z) = \sum_{i=0}^{n-1} \mathbf{w}_i(x,z)R_i(x,z),
\end{equation}
where $\mathbf{w}$ is the apodization window of length $n$. In practice, however, dynamic receive beamforming is utilized wherein the f-number (denoted by f\#) is fixed for the entire image. Therefore, $l$ is defined as the number of crystal elements considered for the reconstruction of each depth ($z$) of the image and is calculated as follows~\cite{4816058}:
\begin{equation} 
	\label{eq:6}
	f\# = z/l.
\end{equation}
After construction of $S$, it is subject to envelope detection and $\log$ compression in order to obtain the final B-mode ultrasound image.
\subsection{Optimal ultrasound PWI}
\label{sec:sec22}
As mentioned in Section~\ref{sec:sec1}, one of the challenges of using deep learning for ultrasound image reconstruction is the lack of ground-truth. In other words, the best output quality in real \textit{in vivo} images is unknown since TRF is unknown. In simulated ultrasound images, however, this problem can be resolved since a complete knowledge of scatterers’ location and amplitude is available. And, researchers use a predefined desired model for the output. For example, in~\cite{8663450}, a despeckled ultrasound image was considered as the output of the image reconstruction from raw data.\par
Herein, we assume that an optimal ultrasound PWI system produces a sharp Gaussian mask as the resulting image of a point target. Indeed, we consider an ideal impulse response (i.e., PSF) for the ultrasound imaging system. This assumption is held for the result of a single plane-wave transmission. And, it is the best possible image quality that an imaging system can achieve.\par 
As for the simulation data, using the proposed PSF as well as a complete knowledge of scatterers’ location and amplitude (i.e., having the TRF), the corresponding optimal image can be acquired using convolution. More specifically, it has been shown that using the first-order Born approximation and assuming weak scattering for soft tissues, the ultrasound images can be modeled as the convolution between a TRF and a PSF~\cite{016173469301500204}. This linear model can be written as follows:
\begin{equation} 
	\label{eq:7}
	S = X \circledast H + N.
\end{equation}
where $S$ is the RF matrix, $X$ is the TRF, $H$ is the impulse response (i.e., the PSF of the imaging system), and $N$ denotes noise. Fortunately, there is no interfering noise in the simulation case. Therefore, our desired simulated image quality (shown in Fig.~\ref{fig:fig5}(c)) for training is created using this method (Eq.~\ref{eq:7}).
\subsection{Network structure}
\label{sec:sec23}
\begin{figure*}[t!]
	\centering
	\centerline{\includegraphics[width=\textwidth]{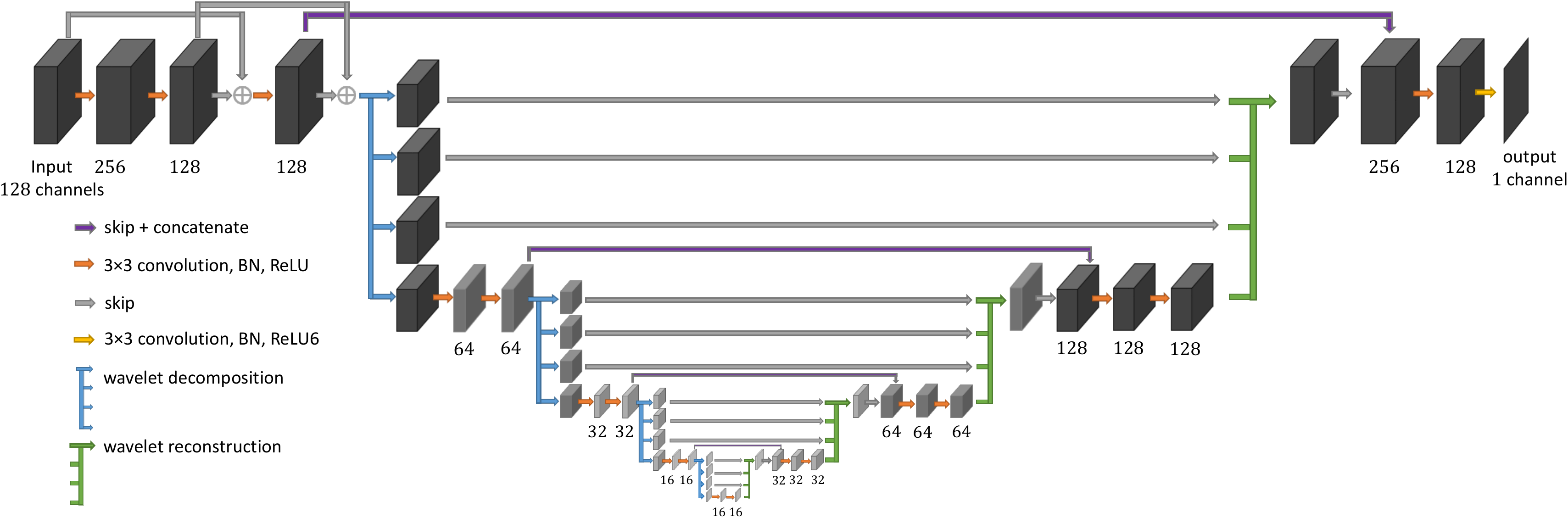}}
	\caption{The structure of the proposed network.}
	\label{fig:fig6}
\end{figure*} 
In order to efficiently learn the mapping function between input-output pairs, the tight frame U-Net~\cite{8332969} structure is modified based on the requirements of our beamforming problem. The proposed network is shown in Fig.~\ref{fig:fig6}. The motivation for the design choices in the proposed structure can be explained as follows. The input data contains all RF matrices of piezoelectric elements reconstructed using Eq.~\ref{eq:4}. This is the rawest possible form of the data that can be used for the proposed beamforming approach in order to make sure that there was not any loss of information in preprocessing steps. As seen in Fig.~\ref{fig:fig6}, there are 3 convolutional layers with skip connections on top of the input to represent the raw data in the new hidden domain. Intuitively, initial layers extract the proper representation of input data for the following encoder-decoder layers while the skip connections provide an alternative path for the gradient. Next, the network contains encoder and decoder parts implemented using wavelet transform. It has been theoretically shown that pooling and unpooling layers in the original structure of U-Net~\cite{2015unet} leads to an overemphasis of the low-frequency components of images due to the duplication of the low-frequency branch~\cite{17M1141771}. In contrast, using wavelet decomposition and reconstruction helps effective recovery of high-frequency content of the image~\cite{8332969}. Herein, Daubechies mother wavelet with 4 vanishing moments is used for  the wavelet transform since it has the most similarity with the collected ultrasound RF signals as compared to other mother wavelets. On one hand, the encoder part of the network successively represents the context of input data as it mainly focus on the approximation branch of wavelet transform. On the other hand, the decoder part successively reconstructs the output from different levels of input's abstraction acquired in encoder layers. The skip connections of encoder-decoder part is used for concatenating the detail branches of wavelet transform in order to keep the high-frequency information.\par
All convolutional layers contain kernels of size 3. Moreover, Batch Normalization (BN) as well as ReLU activation function are used after each convolutional layer. The depth of encoder-decoder part of the network, as well as the number of filters in each layer are selected by trial and error to get the best result on the validation dataset using the lowest possible number of trainable parameters.\par
As the input RF data is in the [-1,1] range and ReLU activation function only works on the positive portion of its input, the first convolutional layer contains the double number of filters to ensure that enough potential for successful mapping of data into the hidden feature space is provided. The network's output is designed to be the ultrasound envelope data. Therefore, the last layer contains a scaled version of ReLU6 activation function in order to make sure that the network output is in the range [0,1]. In total, the network contains 1.5 million trainable parameters.  
\section{Experiments}
\label{sec:sec3}
\subsection{Datasets}
\label{sec:sec31}
\subsubsection{Simulation data}
\label{sec:sec311}
The Field II simulation package~\cite{139123,Jensen96field} is used to simulate ultrasound images. Probe settings are set exactly the same as 128 element linear array transducer L11-4v (Verasonics Inc., Redmond, 240 WA), and the plane-wave dataset is simulated using transmissions at 5.208 MHz with 67\% bandwidth and the sampling frequency of 104.16 MHz. Scatterers with normally distributed amplitudes are uniformly distributed in a 20×45×5~mm\textsuperscript{3} phantom centered at an elevational focus of 2~cm. Scatterer density is set to 60 per resolution cell in order to simulate fully developed speckle. To have enough variety of ground-truth echogenicity maps in the training dataset, scatterers' amplitudes are weighted according to real photographic images. This approach for simulating phantoms with a wide variety of echogenicity
maps was first proposed in~\cite{8663450}. Herein, 512 images are randomly selected from an ImageNet~\cite{ILSVRC15} validation set, and 512 images are randomly selected from the Places2~\cite{7968387} test set. Each image is converted to grayscale and cropped into a 200 pixels × 450 pixels patch, and then mapped onto the lateral and
axial extents of the phantom (20 mm × 45 mm) to be used as the ground-truth echogenicity map. Therefore, the scatterers' amplitudes are weighted based on the pixel intensities of images according to their positions via bilinear interpolation. Each image is simulated using an independent realization of scatterers. In addition to 1024 phantom images simulated with the explained method, 150 ultrasound images containing a randomly selected number of point targets that are randomly distributed over an anechoic background are also simulated to provide the network with images of sparsely distributed scatterers. In total, the simulation dataset contains 1174 images.\par
In order to evaluate the performance of the proposed method, simulation ultrasound dataset that is publicly available through UltraSound ToolBox (USTB)~\cite{8092389} is used. More specifically, the simulation dataset contains one image containing point targets distributed vertically and horizontally over an anechoic background designed to measure the spatial resolution, and one image containing anechoic cysts distributed vertically and horizontally over fully developed
speckle to measure the contrast. It has to be mentioned that our simulation settings are exactly the same in this benchmark dataset.
\subsubsection{\textit{In vivo} data}
\label{sec:sec312}
\textit{In vivo} ultrasound dataset, publicly available through USTB, is also used for performance evaluation. This dataset contains cross-sectional and longitudinal views of the carotid artery of a volunteer. Plane-wave images were collected using a Verasonics Vantage 256 research scanner and a L11-4v probe (Verasonics Inc., Redmond, 240 WA).
\subsection{FCNN training}
\label{sec:sec32}
Only our simulation dataset is used for training, leaving data provided through USTB for testing. More specifically, 90\% of our images are used for training and 10\% for validation. The network's input consists of 128 channels corresponding to RF matrices of piezoelectric elements reconstructed using Eq.~\ref{eq:4}. Each element gives an RF matrix of size 2688×384, and is broken into two patches of size 1344×384 due to memory limitations during training. The network's output is the corresponding envelope image of the desired TRF created with the proposed approach outlined in Section~\ref{sec:sec22}. Therefore, we have 2113 input-output pairs for training and 235 for validation.\par
The network is trained in four different steps from coarse (mean square error) to fine ($\ell_{0.2}$) loss functions. In each step, one distinct loss function is used. More specifically, the training is started with Mean Square Error (MSE) loss function calculated as follows:
\begin{equation} 
	\label{eq:10}
	MSE=\frac{1}{m} \sum_{i=1}^m (y_i-\widehat{y_i})^2
\end{equation}
where $m$ is the batch size of data in each iteration, $y$ and $\widehat{y}$ are the desired output and the estimated network's output, respectively. A learning rate of 10\textsuperscript{-3} is used in this step for 25 epochs which produces blurry results. This result can be considered as a rough approximation of the final output in which fine details are not present. In the second step, Mean Absolute Error (MAE), also known as the $\ell_{1}$, loss function calculated as following:
\begin{equation} 
	\label{eq:11}
	MAE=\frac{1}{m} \sum_{i=1}^m \mid y_i-\widehat{y_i} \mid
\end{equation}
with a learning rate of 5×10\textsuperscript{-4} is used for 50 epochs. Compared to the previous step, the network is trained to reconstruct more details of the output. Afterward, the network is trained with $\ell_{0.4}$ loss function, with a learning rate of 10\textsuperscript{-4} for 25 epochs.
\begin{equation} 
	\label{eq:12}
	\ell_{0.4}=\frac{1}{m} \sum_{i=1}^m \mid y_i-\widehat{y_i} \mid ^{0.4}
\end{equation}
Finally, $\ell_{0.2}$ loss function with a learning rate of 10\textsuperscript{-5} is used for the last 25 epochs. 
\begin{equation} 
	\label{eq:13}
	\ell_{0.2}=\frac{1}{m} \sum_{i=1}^m \mid y_i-\widehat{y_i} \mid ^{0.2}
\end{equation}
During the last two steps of training, fine details of the output are reconstructed. As mentioned above, the learning rate is successively decreased in each step to prevent a major distortion on trained parameters. Step-by-step training also helps training error reduces faster as opposed to starting with $\ell_{0.2}$ loss function from scratch. That is because in the initial stage of training, $\ell_{0.2}$ loss is very large. Therefore, the learning rate has to be selected as a very small number which renders training very slow. Otherwise, the training procedure does not converge, as theoretically outlined in curriculum learning~\cite{10.1145/1553374.1553380}.\par
In each step, the training procedure is stopped once the error curve becomes flat. The overfitting problem on the training set has not been noticed since the size of training and validation sets are very large. The proposed network is implemented using the PyTorch library and training was done with an Nvidia Titan Xp GPU. The batch size is 2, and AdamW~\cite{loshchilov2017decoupled} optimizer with $\beta_1 = 0.9$ and $\beta_2 =0.999$ is used in all steps. As for the wavelet transform in the proposed network, the PyWavelets library is used, which provides the possibility of running wavelet transform on the GPU. 
\subsection{GAN training}
\label{sec:sec33} 
As mentioned before, the goal is to estimate a nonlinear function that maps the RF channel data to the proposed ideal image. This goal can be fulfilled using the FCNN explained in the previous section. However, FCNN training needs an explicit differentiable objective function to specify the quality of the network's output. On the contrary, Generative Adversarial Network (GAN) provides the possibility of learning an objective function appropriate for satisfying a specific task. More specifically, GAN consists of generator and discriminator networks competing with each other. The generator network is the mapping function which estimates the desired output. The discriminator network works as a loss function which specifies whether the estimated output is real or fake.\par      
GAN training is a min-max game between the generator and the discriminator:
\begin{equation}
	\label{eq:9} 
	\begin{split}
		\min_{G}\max_{D}V(D,G)=E_{y\sim{p_{data}(y)}}\left[log\ D(y)\right]\\+E_{x\sim{p_{x}(x)}}\left[log(1- D(G(x))\right]
	\end{split}
\end{equation}
where $D$ and $G$ are the discriminator and generator, respectively. $y$ and $x$ are the desired output and input, respectively. ${E}$ denotes the expected value, and $p(\cdot)$ denotes probability of the enclosed parameter. More details regarding GAN training can be found in~\cite{goodfellow2014generative}.
\subsection{Evaluation metrics}
\label{sec:sec34}
The reconstructed images are evaluated in terms of resolution and contrast. As for resolution, the Full-Width at Half-Maximum (FWHM) index of point targets is calculated in both axial and lateral directions. As for the contrast, a variety of indexes are reported. Speckle Signal-to-Noise Ratio (SSNR)~\cite{5953993} as well as Contrast Ratio (CR)~\cite{SMITH1985467} are calculated as follows:
\begin{equation} 
	\label{eq:1}
	SSNR = \frac{\mu_{B}}{\sigma_{B}},
\end{equation} 
\begin{equation} 
	\label{eq:2}
	CR=20log_{10}(\frac{\mu_{ROI}}{\mu_{B}}),
\end{equation} 
where $\mu_{B}$ and $\mu_{ROI}$ are the mean of envelope image (before log compression) over the background and the region of interest, respectively. $\sigma_{B}$ is the standard deviation of the envelope image over the background region.\par
The other assessed contrast index  is a generalization of Contrast-to-Noise Ratio (CNR) definition, which was recently introduced~\cite{8918059}. It is a robust index of contrast and lesion detectability against dynamic range alterations and called generalized CNR (gCNR). gCNR is calculated as follows:
\begin{equation} 
	\label{eq:3}
	gCNR = 1- \int_{-\infty}^{\infty} min\left\{p_{ROI}(x),p_B(x)\right\}dx
\end{equation} 
where $p_B(x)$ and $p_{ROI}(x)$ are the histograms of pixels measured in the region of interest and background, respectively. gCNR determines how much the intensity distributions of two regions are overlapped regardless of grayscale intensity transformations. Lower distributions overlap leads to higher gCNR values. gCNR is equal to its maximum value of 1 when the two distributions are independent~\cite{8918059}.
\section{Results}
\label{sec:sec4}
Herein, all the presented results correspond to the test set, which have not been used during training or validation process. More specifically, the trained network, with the best results on the validation set, is evaluated on two test sets of simulation and \textit{in vivo} data taken from USTB~\cite{8092389}. It has to be emphasized that the test experiments are completely blind and all results are produced without any further fine-tuning.\par
The proposed method is implemented using the proposed Fully Convolutional Neural Network (FCNN) as well as GAN. FCNN is trained based on the method explained in Section~\ref{sec:sec32}. GAN is trained using adversarial training in which an extra discriminator network provides an additional loss function.\par
The result of the proposed method is compared with other approaches including DAS beamforming on the single $0^{\circ}$ plane-wave as well as 75 plane-waves (the result of CPWC), Minimum Variance (MV) beamforming~\cite{5278437}, and our previously published method, referred to as MobileNetV2 method~\cite{9251565}, which was ranked 1 (in terms of beamforming quality) in the CUBDL 2020  Challenge~\cite{9251434}. 
\subsection{Simulation data}
\label{sec:sec41}
\subsubsection{Resolution distortion}
\label{sec:sec411}
\begin{figure*}[t!]
	\centering
	\centerline{\includegraphics[width=\textwidth]{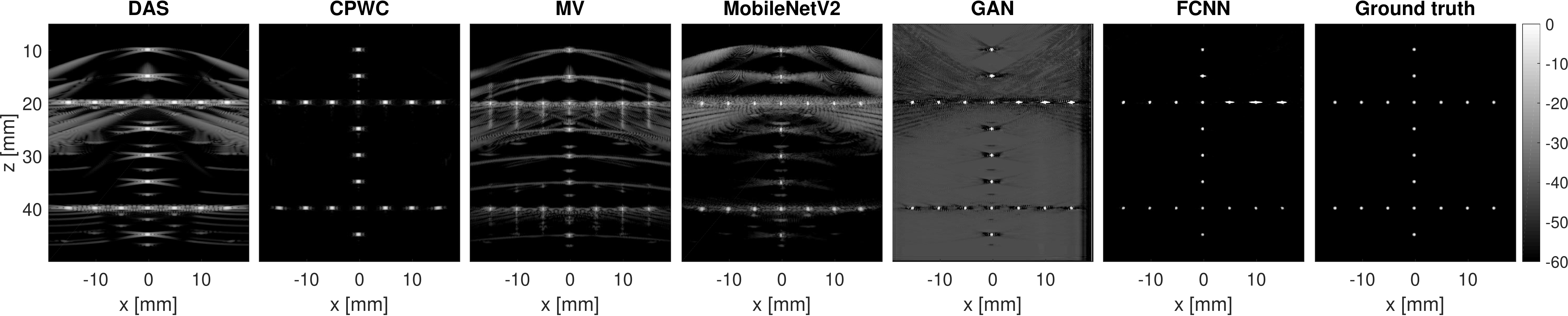}}
	\caption{Results on simulation resolution data. Point targets are distributed vertically and horizontally over an anechoic background. CPWC is obtained from 75 steered insonifications. All other results are from a single insonification.}
	\label{fig:fig1}
\end{figure*} 
Fig.~\ref{fig:fig1} illustrates the results of different methods on the simulation resolution test image. Although the MV beamformer, as well as MobileNetV2, produce high lateral resolutions, they mainly suffer from additional artifacts. In the CPWC method, these additional artifacts are completely removed by averaging among several insonifications at different angles. It is worth noting that the proposed ground-truth (produced with the method explained in Section~\ref{sec:sec22}) offers the best quality without any artifact with excellent lateral and axial resolutions, even better than that of optimal CPWC. The proposed beamforming approach using FCNN produces an image quality similar to the ground-truth while adversarial training does not provide good result. FCNN outperforms GAN because the adversarial training produces biased outputs with additional artifacts in the two lateral ends of the output. This reduction in quality might be due to the nonuniform ultarsound image quality in the lateral direction which affects the corresponding adversarial loss. It is important to note that the gain in quality in CPWC comes at the expense of a considerable X 75 drop in the framerate.\par
Table~\ref{table:1} includes a quantitative comparison of different methods in terms of axial and lateral resolution.  As mentioned, the proposed FCNN is of worse lateral resolution compared to MV and MobileNetV2. However, as seen in Fig.~\ref{fig:fig1}, it does not have any side-lobe artifacts. Furthermore, the proposed approach interestingly improves the axial resolution, while other beamforming methods do not have any effect on the axial resolution. It is important to note that while advanced beamforming techniques can be used for improving lateral resolution, options for improving the axial resolution are very limited, with the most common solution as increasing the center frequency. Increasing this frequency comes at the expense of a loss in penetration depth due to the rapidly increasing attenuation with frequency.\\   
\begin{table}[b!]
	\caption{Quantitative results in terms of resolution and contrast indexes for simulation test experiments.}
	\label{table:1}
	\centering
	\setlength{\tabcolsep}{2.5pt}
	\scriptsize
	\begin{tabular}{c c c c c c c c c c c c c c c c c}
		\specialrule{.15em}{0em}{.2em}
		dataset & SR && SC  \\ [.2em] 
		\specialrule{.05em}{0em}{.2em} 
		index & FWHM\textsubscript{A} FWHM\textsubscript{L} & SSNR & CR & gCNR \\ [.2em] 
		\specialrule{.05em}{0em}{.2em} 
		\makecell{DAS \\ CPWC \\ MV \\ MobileNetV2 \\ GAN \\FCNN \\ Ground-truth} & \makecell{0.4 \\ 0.39 \\ 0.41 \\ 0.42 \\ 0.29 \\ 0.25 \\ 0.23} \makecell{0.82 \\ 0.56 \\ 0.1 \\ 0.27 \\ 0.4 \\ 0.28 \\ 0.22} &    \makecell{1.34 \\ 1.21 \\ 1.16 \\ 1.05 \\ 1.4 \\ 2.31 \\ 1.8} & \makecell{-15.15 \\ -31.44 \\ -21.15 \\ -17.15 \\ -29.32 \\ -39.44 \\ -47.64}& \makecell{0.74 \\ 0.97 \\ 0.82 \\ 0.66 \\ 0.94 \\ 0.99 \\ 1}  \\ [.2em] 
		\specialrule{.05em}{0em}{.2em}
	\end{tabular}
\end{table}
\subsubsection{Contrast speckle}
\label{sec:sec412}
\begin{figure*}[t!]
	\centering
	\centerline{\includegraphics[width=\textwidth]{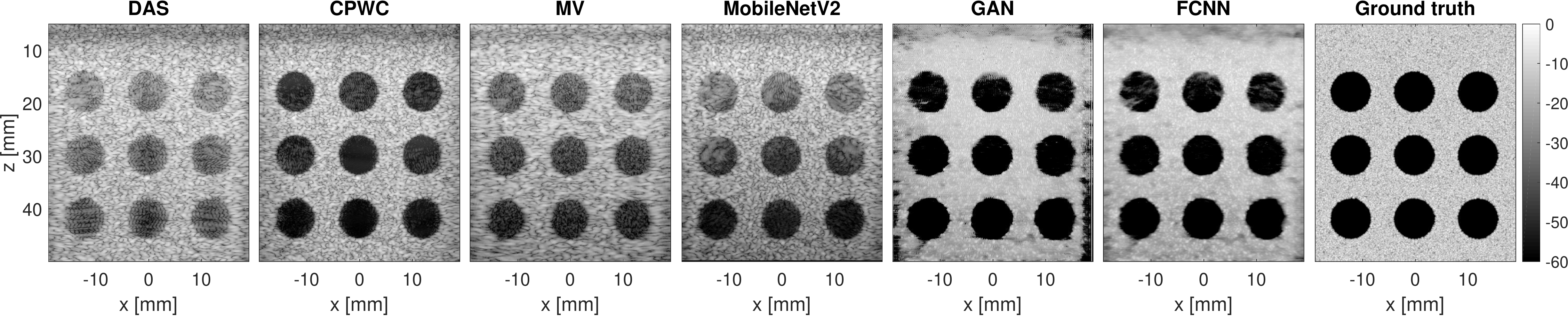}}
	\caption{Results on simulation contrast speckle data. Anechoic cysts are distributed vertically and horizontally over fully developed speckle. CPWC is obtained from 75 steered insonifications. All other results are from a single insonification.}
	\label{fig:fig2}
\end{figure*}

Fig.~\ref{fig:fig2} shows the output of different methods for the simulation contrast test image. The quality of cyst regions is limited in DAS, MV, and MobileNetV2 methods mainly due to having unfocused transmissions degrading the cyst contrast. This problem is resolved in CPWC, and a noticeable improvement of contrast is visible in the cyst regions. The proposed ground-truth image, however, can be considered as the ideal ultrasound image quality that can be acquired since it gives perfect gCNR (Table~\ref{table:1}). As shown in Fig.~\ref{fig:fig2}, the result of the proposed method using FCNN depicts the most similarity to the desired image. This point is also quantitatively confirmed in Table~\ref{table:1}. 
\subsection{\textit{In vivo} data}
\label{sec:sec42}
\begin{figure*}[t!]
	\centering
	\centerline{\includegraphics[width=\textwidth]{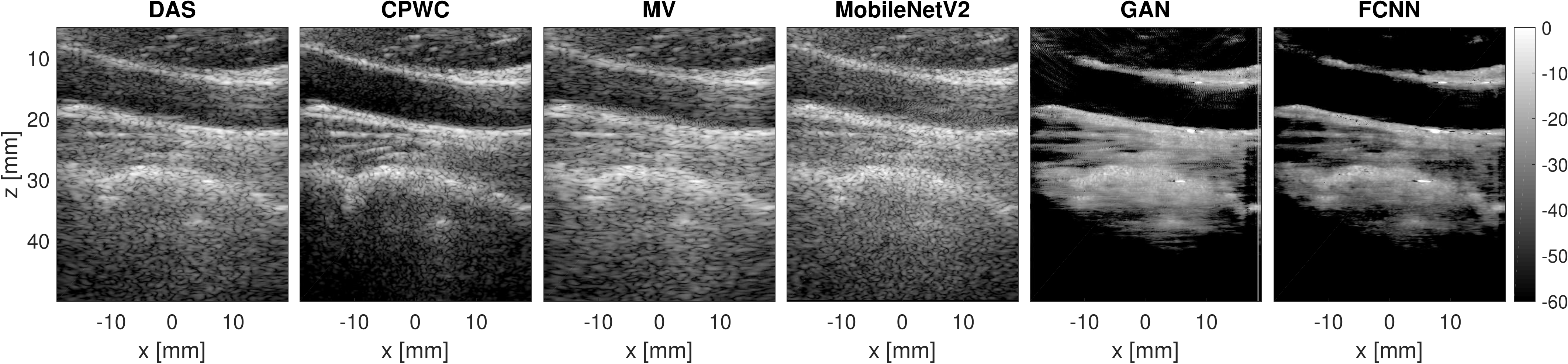}}
	\caption{Results on \textit{in vivo} data. This image shows a longitudinal view of the carotid artery of a volunteer. CPWC is obtained from 75 steered insonifications. All other results are from a single insonification.}
	\label{fig:fig3}
\end{figure*}

\begin{figure*}[b!]
	\centering
	\centerline{\includegraphics[width=\textwidth]{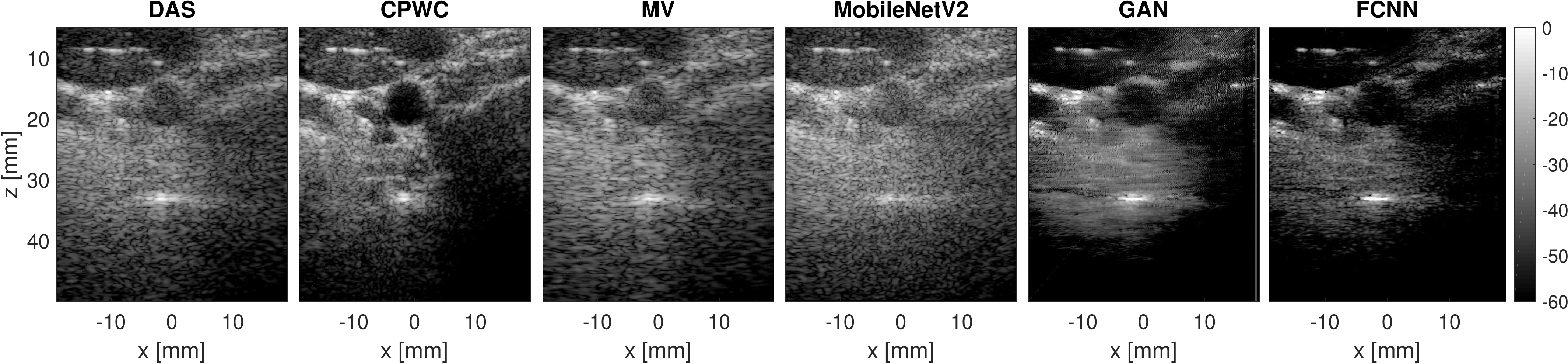}}
	\caption{Results on \textit{in vivo} data. This image shows a cross-sectional view of the carotid artery of a volunteer. CPWC is obtained from 75 steered insonifications. All other results are from a single insonification }
	\label{fig:fig4}
\end{figure*}

The results on the \textit{in vivo} test datasets are shown in Fig.~\ref{fig:fig3} and Fig.~\ref{fig:fig4}. Herein, we do not have the ground-truth anymore. However, we expect the artery to be fully dark, and the brightness within the artery in DAS, MV, and MobileNetV2 is a result of the clutter created by diffuse reverberation from shallow layers~\cite{9251465,5750097,DAHL2014714}.
As shown in both cross-sectional and longitudinal views, the proposed method based on FCNN depicts the carotid artery without any clutter artifact, while the GAN implementation is not of similar quality. Other approaches including CPWC mainly suffer from the clutter artifact in the carotid artery regions. The \textit{in vivo} results worth noting since there is always a domain shift between simulation and real experimental data due to several factors not modelled in simulation such as nonlinear acoustics, phase aberration, and multipath scattering. This domain shift adversely affects the network’s performance. However, our proposed method maintains good performance in the new domain without any extra fine-tuning. 
\section{Discussions}
\label{sec:sec5}
The robustness of our proposed beamforming approach in the real test experiment mainly comes from the following reasons. First, using real photographic images as the ground-truth echogenicity map of simulation phantoms provides the network with enough variety of textures, contrasts, and objects with different geometries in the training phase. It also makes the simulation of a large number of training and validation data possible, which is important to prevent overfitting. Furthermore, using the simulation settings similar to the real experimental imaging settings of \textit{in vivo} test data minimizes the unwanted domain shift between training and test datasets. The performance of the proposed FCNN may be further improved by fine-tuning on wire phantoms wherein the exact locations of wires is known. \par
The proposed method for simulating the training ground-truth provides images with the best possible quality in ultrasound imaging. As seen in the simulation resolution results (Fig.~\ref{fig:fig1}), the proposed ground-truth provides excellent lateral as well as axial resolution. It is also free from off-axis scattering. As for the simulation contrast results (Fig.~\ref{fig:fig2}), the ground-truth image is of perfect contrast while there is no trace of speckle noise.\par
By considering RF matrices of all piezoelectric elements as the network's input, we ensure that all existing information of the backscattered signals are available in the input domain, and the rawest possible format is provided for the network. It has to be mentioned that if we do not apply the propagation delays to the element's outputs, the task is not tractable for the network, and training does not converge. As for the network output, there are two possibilities including the definition of ground-truth in the format of RF or envelope data. Herein, the second one is used since our goal was to recover the envelope data. Since some applications need the RF data, we plan to extend the proposed method to define the desired PSF in the RF format in the future.\par
The achieved increase in quality of plane-wave ultrasound imaging is of crucial importance in practice since there is no longer a need to transmit several insonifications at different angles to achieve the optimal quality of CPWC. In other words, the proposed method goes beyond the classical limitations and can be considered as a solution to the intrinsic trade-off between framerate and image quality of plane-wave imaging. As shown in the results, the proposed method also provides improvements in the axial direction, which otherwise is usually achieved with higher center frequencies. As such, the proposed method can also be considered as a potential solution to the intrinsic trade-off between penetration depth and axial resolution. Further validation experiments are necessary to benchmark the efficacy of the proposed method in diagnosis and image-guided interventions.
\section{CONCLUSIONS}
\label{sec:sec6}
A reduction in the frame-rate is the main challenge associated with CPWC. Herein, the proposed beamforming approach works as a nonlinear mapping function from the input space (ultrasound RF channel data) to the ideal output image. As shown in the results, an FCNN with the proposed step-by-step method of training has been adapted to achieve the quality of ideal ultrasound images without any loss in frame-rate. The experiments confirm that the proposed method reconstructs images with a high quality in terms of resolution and contrast, while it also preserves the performance on the \textit{in vivo} datasets.
\section{Declaration of Competing Interest}  
All authors declare that the manuscript is not affected by any conflict of interests, financial and personal relationships with other people or organisations that could inappropriately influence this work.
\section*{Acknowledgment}
\label{sec:sec7}
We acknowledge the support of the Natural Sciences and Engineering
Research Council of Canada (NSERC) RGPIN-2020-04612. We would like to thank NVIDIA for the donation of the Titan Xp GPU.
\bibliography{refs}
\end{document}